\documentclass{report}

\usepackage{ohm-ngc}

\usepackage{epsfig}
\usepackage {graphicx}
\usepackage{amsmath}

\begin{document}
\pagestyle{headings}


\title{Exact Cover with light}

\author{Mihai OLTEAN, Oana MUNTEAN}
{Department of Computer Science,\\
Faculty of Mathematics and Computer Science,\\
Babe\c s-Bolyai University, Kog\u alniceanu 1,\\
Cluj-Napoca, 400084, Romania.\\
}
\E-mail{moltean@cs.ubbcluj.ro\\}
\E-mail{www.cs.ubbcluj.ro/$\sim$moltean}


\begin{abstract}

We suggest a new optical solution for solving the YES/NO version of the Exact Cover problem by using the massive parallelism of light. The idea is to build an optical device which can generate all possible solutions of the problem and then to pick the correct one. In our case the device has a graph-like representation and the light is traversing it by following the routes given by the connections between nodes. The nodes are connected by arcs in a special way which lets us to generate all possible covers (exact or not) of the given set. For selecting the correct solution we assign to each item, from the set to be covered, a special integer number. These numbers will actually represent delays induced to light when it passes through arcs. The solution is represented as a subray arriving at a certain moment in the destination node. This will tell us if an exact cover does exist or not.

\end{abstract}


\section{Introduction}

In this paper we propose an unconventional optical device, which uses light, for solving the Exact Cover (XC) problem \cite{garey}. This problem can be simply stated as follows: Given a set $U$ and $C$ a set of subsets of $U$, the problem asks to find if there is a subset $S$ of $C$ such that each element of $U$ appears once in $S$. 

We have decomposed the problem in 2 subproblems: generating all subsets of $C$ and then selecting the correct one.

For the first step we have designed a light-based device which has a graph-like structure. Each arc can represent either an element of $C$ or can be a skipping arc. An arc will actually delay the signal (light) which passes through by a certain amount of time. The nodes are connected by arcs in such way all possible subsets of $C$ are generated. 

For the second part we have assigned, to each item from $U$, a special positive number such that sum of all numbers assigned to $U$ is not equal to any other combination of numbers assigned to items from $U$. The delay induced by an arc is actually equal to the sum of numbers belonging to the element from $C$ attached to that arc.

Initially a light ray is sent to the $Start$ node. In each node the light is divided into 2 subrays. Each arc delays the ray by an amount of time equal to the number assign to it. Let us denote by $B$ the sum of the numbers assigned to items from $U$. At the $Destination$ node we will check if there is a ray arriving at the moment equal to $B$ (plus some constants introduced by the system). If there is such signal it means that we can have an exact cover for $U$. Otherwise $U$ does not have an exact cover. This is guaranteed by the special numbers assigned to each item from $U$.

The paper is organized as follows: Related work in the field of optical computing is briefly overviewed in section \ref{related}. The Exact Cover problem is described in section \ref{subsetsum}. Useful properties of light are discussed in section \ref{useful}. The proposed device is presented in section \ref{proposed}. We describe both the way in which all possible subsets of a given set are generated (see section \ref{basic}) and we also show how to detect the set covering $U$ exactly (see section \ref{labeling}). The way in which the proposed device works is given in section \ref{howorks}. A list of components required by the proposed device is given in section \ref{hard}. Weaknesses of our device are discussed in sections \ref{precision}-\ref{tech}. Other suggestions for improving the device are given in section \ref{speed_reduce}. Further work directions are suggested in section \ref{further}.

\section{Related work}
\label{related}

Using light instead of electric power for performing computations is not a new idea. Optical Character Recognition (OCR) machines \cite{wiki} were one of the first modern devices which are based on light for solving a difficult problem. Later, various researchers have shown how light can solve problems faster than modern computers. An example is the $n$-point discrete Fourier transform computation which can be performed in only unit time \cite{goodman,reif}.

Since each such gate has two input signals and only one output signal, such architectures are fundamentally dissipative in information and energy. Their serial nature also induces a latency in the processing time. 

In \cite{hardy} was presented a new, principally non-dissipative digital logic architecture which involves a distributed and parallel input scheme where logical functions are evaluated at the speed of light. The system is based on digital logic vectors rather than the Boolean scalars of electronic logic. This new logic paradigm was specially developed with optical implementation in mind. 

An important practical step was made by Intel researchers who have developed the first continuous wave all-silicon laser using a physical property called the Raman Effect \cite{Faist,Paniccia,rong1,rong2}. The device could lead to such practical applications as optical amplifiers, lasers, wavelength converters, and new kinds of lossless optical devices.

Another solution comes from Lenslet \cite{lenslet} which has created a very fast processor for vector-matrix multiplications.  This processor can perform up to 8000 Giga Multiple-Accumulate instructions per second. Lenslet technology has already been applied to data analysis using $k-$mean algorithm and video compression.

In \cite{Schultes} the idea of sorting by using some properties of light is introduced. The method called Rainbow Sort is based on the physical concepts of refraction and dispersion. It is inspired by the observation that light that traverses
a prism is sorted by wavelength. For implementing the Rainbow Sort one need to perform the following steps:

\begin{itemize}

\item{encode multiple wavelengths (representing the numbers to be sorted) into a light ray,}

\item{send the ray through a prism which will split the ray into $n$ monochromatic rays that are sorted by wavelength,}

\item{read the output by using a special detector that receives the incoming rays.}

\end{itemize}

A stable version of the Rainbow Sort is proposed in \cite{Murphy}.

Naughton (et al.) proposed and investigated \cite{Naughton,woods} a model called the continuous space machine which operates in discrete time-steps over a number of two-dimensional complex-valued images of constant size and arbitrary spatial resolution. The (constant time) operations on images include Fourier transformation, multiplication, addition, thresholding, copying and scaling.

An optical solution for solving the traveling salesman problem (TSP) was proposed in \cite{Shaked}. The power of optics in this method was done by using a fast matrix-vector multiplication between a binary matrix, representing all feasible TSP tours, and a gray-scale vector, representing the weights among the TSP cities. The multiplication was performed optically by using an optical correlator. To synthesize the initial binary matrix representing all feasible tours, an efficient algorithm was provided. However, since the number of all tours is exponential the method is difficult to be implemented even for small instances.

An optical system which finds solutions to the 6-city TSP using a Kohonen-type network was proposed in \cite{Collings}. The system shows robustness with regard to the light intensity fluctuations and weight discretization which have been simulated. Using these heuristic methods, a relatively large number of TSP cities can be handled.

A system which solves the Hamiltonian path problem (HPP) \cite{garey} by using light and its properties has been proposed in \cite{oltean_uc}. The device has the same structure as the graph where the solution is to be found. The light is delayed within nodes, whereas the delays introduced by arcs are constants. Because the problem asks that each node has to be visited exactly once, a special delaying system was designed. At the $Destination$ node, a ray which has visited each node exactly once has been searched. This is very easy due to the special properties of the delaying system.

A similar idea was used in \cite{haist} for solving the TSP problem. The device uses white light interferometry for find the shortest TSP path.

\section{The Exact Cover problem}
\label{subsetsum}

The description of the Exact Cover problem \cite{garey} is the following:

Given a set $U = \{u_1, u_2, ..., u_n\}$ of elements and a set $C$ of subsets of $U$, an exact cover is a subset $S$ of $C$ such that every element in $U$ is contained in exactly one set in $S$.

Our purpose is to solve the YES/NO decision problem. This means that we are not interested to compute the actual subset $S$. Rather, we are interested in finding whether a solution does exist or not.

The problem belongs to the class of NP-complete problems \cite{garey}. No polynomial time algorithm is known for it. 

In what follows we denote by $m$ the cardinal of $C$.

\subsection{Example}
\label{xc_example}

$U = \{u_1, u_2, ..., u_5\}$

$C = \{C_1, C_2, C_3, C_4, C_5, C_6\}$

$C_1 = \{u_1, u_3\}$

$C_2 = \{u_1, u_2, u_5\}$

$C_3 = \{u_2\}$

$C_4 = \{u_1, u_4, u_5\}$

$C_5 = \{u_2, u_3, u_4\}$

$C_6 = \{u_4, u_5\}$\\

An exact cover for $U$ is :\\

\[
S = \{C_1, C_3, C_6\}.
\]

We are not thinking to $U$ as a set of numbers. Rather we adopt a more general definition: $U$ is a set of items. We do this in order to avoid confusion which might appear because, later in this paper (see section \ref{labeling}), we will attach positive integer numbers to each item in $U$. Those numbers have some special properties and they should not be confounded with the actual value of the items $u_i$ (in the case that these values are numerical).

\section{Why light is good for our purpose ?}

In this section we argue why light can be used as possible medium for the implementation. We also describe the operations which are performed within our device.

\subsection{Useful properties of light}
\label{useful}

Our idea is based on two properties of light:

\begin{itemize}

\item{The speed of light has a limit. The value of the limit is not very important at this stage of explanation. The speed will become important when we will try to measure the moment when rays arrive at the $Destination$ node (see section \ref{precision}). What is important now is the fact that we can delay the ray by forcing it to pass through an optical fiber cable of a certain length.}

\item{The ray can be easily divided into multiple rays of smaller intensity/power. Beam-splitters are used for this operation.}

\end{itemize}

\subsection{Operations performed within our device}
\label{operations}

The proposed device has a graph like structure. Generally speaking one operation is performed when a ray passes through a node and one operation is performed when a ray passes through an arc.

\begin{itemize}

\item{When passing through an arc the light ray is delayed by the amount of time assigned to that arc.}

\item{When the ray is passing through a node it is divided into a number of rays equal to the external degree of that node. Each obtained ray is directed toward one of the nodes connected to the current node.}

\end{itemize}

\section{The proposed device}
\label{proposed}

We have divided our problem into 2 small subproblems:

\begin{itemize}

\item{to generate all possible subsets of $C$. We show how our device is capable of performing this operation in section \ref{basic}.}

\item{to find if there is a subset of $C$ which contains each element of $U$ exactly once. This is possible due to the special properties of the numbers assigned to each item from $U$ (see section \ref{labeling}).}

\end{itemize}

\subsection{Generating all subsets of $C$}
\label{basic}

First of all we assign to each item $u_i$ from $U$ a positive number ($d_i$). The way in which these numbers are constructed is described in section \ref{labeling}. We will use these numbers when we will compute the moment when a particular ray has arrived at the output of our device. The sum of all these numbers is denoted by $B$.

In this section we want to design a device that will generate all possible subsets of $C$.

The first idea for our device was that elements from the given set $C$ represent the delays induced to the signals (light) that passes through our device. For instance, if elements $C_1$, $C_3$ and $C_6$ generate the expected exact cover, then the total delay of the signal should be $delay(C_1) + delay(C_3) + delay(C_6)$. By $delay(C_i)$ we denote the delay introduced by $C_i$ (which is a subset of $U$). The $delay(C_i)$ is computed as sum of delays induced by each item (from $U$) belonging to $C_i$.

If using light we can easily induce some delays by forcing the ray to pass through an optical cable of given length.

\begin{figure}[htbp]
\centerline{\includegraphics[width=5.34in,height=1.69in]{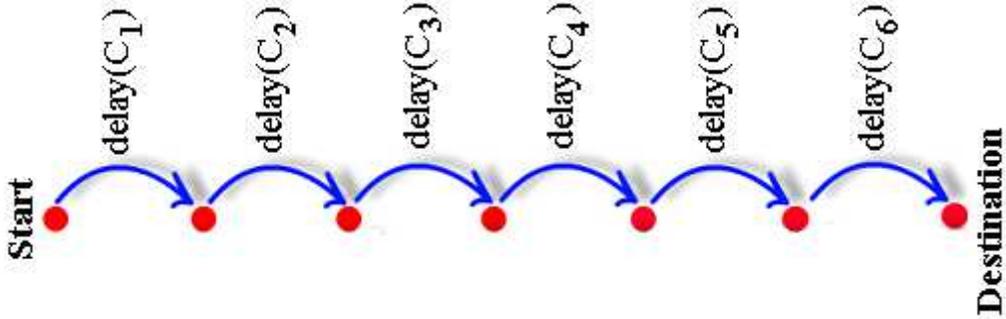}}
\caption{First version of our device. Each arc delays the ray by the amount of time written on it. Note that this device is not complete because it cannot generate all possible subsets of $C$.}
\label{fig_device1}
\end{figure}

This is why we have designed our device as a direct graph. Arcs, which are implemented by using optical cables, are labeled with numbers representing delays of subsets $C_i (1 \le i \le m)$. Each subset is assigned to exactly one arc and there are no two arcs having assigned the same subset. There are $m + 1$ nodes connected by $m$ arcs. At this moment of explanation we have a linear graph as the one shown in Figure \ref{fig_device1}. 

However, this is not enough because we also need an mechanism for skipping an arc encoding an element of $C$. Only in this way we may generate all possible subsets of $C$. 

\begin{figure}[htbp]
\centerline{\includegraphics[width=5.27in,height=1.69in]{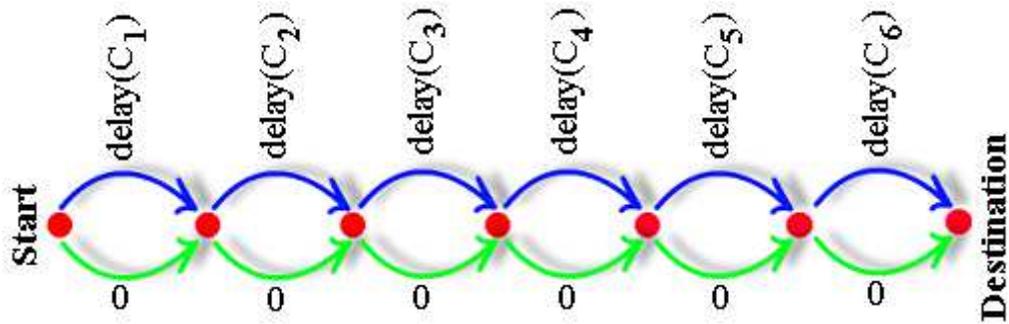}}
\caption{Second version of our device. Each subset of $C$ is generated, but this device cannot be implemented in practice because we cannot have cables of length 0.}
\label{fig_device2}
\end{figure}

A possible way for achieving this is to add an extra arc (of length 0) between any pair of consecutive nodes. Such device is depicted in Figure \ref{fig_device2}. A light ray sent to $Start$ node will have the possibility to either traverse a given arc (from the upper part of figure) or to skip it (by traversing the arc of length 0 from the bottom of figure).

In each node (but the last one) we place a beam-splitter which will split a ray into 2 subrays of smaller intensity. 

The device will generate all possible subsets of $C$. Each subset will delay one of the ray by an amount of time equal to the sum of the lengths of the arcs in that path.

There is a problem here: even if theoretically we could have arcs of length 0, we cannot have cables of length 0 in practice. For avoiding this problem we have multiple solutions. The first one was to use very short cables (let's say of length $\epsilon$) for arcs which are supposed to have length 0. However, there is another problem here: we could obtain for instance the sum $B$ written as $B = a_1 + 3*\epsilon$. Even if there is no subset of sum $B$, still there will be possible to have a signal at moment $B$ due to the situation presented above.

\begin{figure}[htbp]
\centerline{\includegraphics[width=5.32in,height=1.91in]{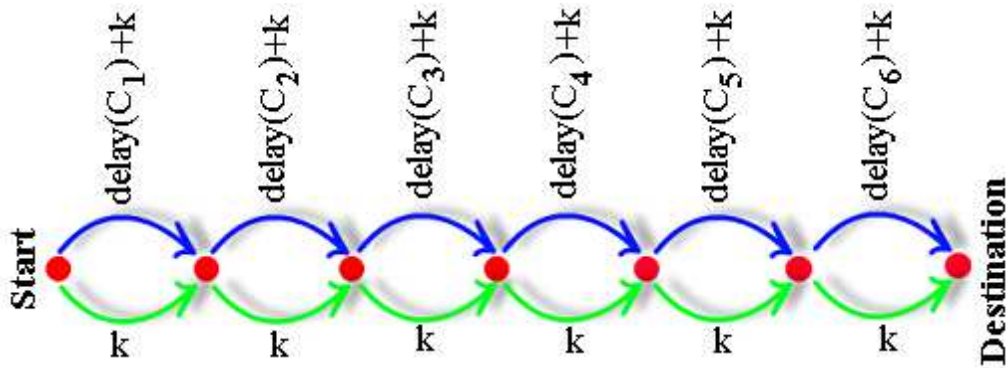}}
\caption{A schematic representation of the device used for solving an instance with 6 elements in $C$. On each arc we have depicted its length. There are $n$ cables of length $k$ and $m$ cables of length $delay(C_i) + k$ ($1 \le i \le m$). This device does generate all possible subsets of $C$ and it can be implemented in practice}
\label{fig_device3}
\end{figure}

For avoiding this situation we have added a constant $k$ to the length of each cable. The schematic view of this device is depicted in Figure \ref{fig_device3}.

We can see that each path from $Start$ to $Destination$ contains exactly $m$ time value $k$. Thus, at the destination we will not wait anymore at moment $B$. Instead we will wait for a solution at moment $B+m*k$ since all subsets will have the constant $m*k$ added.

In Figures \ref{fig_device1}, \ref{fig_device2} and \ref{fig_device3} we have not depicted the items $u_i$ from each $C_k$ ($1 \le k \le m$). The delays introduced by each $u_i$ (see section \ref{labeling}) are actually in the delays of all $C_k$ containing $u_i$.

This trick has solved one of the problems related to our device. It shows how to generate all possible subsets of $C$. However, this is not enough. Another requirement of the problem is that each element of $U$ to be represented once in the exact cover solution. For solving this problem we assign to each element in $U$ a special integer value. These values will be chosen in such way will be possible to easily identify the exact cover solution at the $Destination$ node.

In the next section we show how to decide if is there a subset of $C$ which contains each element of $U$ exactly once.

\subsection{Labeling system}
\label{labeling}

We start by a simple definition about what we mean by a ray passing through an item of $U$.

We say that a ray has visited (passed through) a certain element $u_i$ of $U$ if that ray has passed through an arc representing one of the subsets of $C_k$ which contains element $u_i$. Please note that $C_k$ can contain more elements from $U$ and if one of them is visited by a ray it means that all of them are visited by that ray since the light has no chance of avoiding parts of an arc once it has started to traverse it. 

Take for instance set $C$ from our example (section \ref{xc_example}). If the arc encoding $C_4$ is visited by a light ray it means that all items from that set (namely $u_1$, $u_4$ and  $u_5$) are visited.

At the $Destination$ node we will wait for a particular ray which has visited each element of $U$ exactly once. This is why we need to find a way to label that particular ray so that it could be easily identified.

Recall from section \ref{basic} that the rays passing through an arc are marked by delaying them with a certain amount of time. This delay can be easily obtained by forcing the rays to pass through an optical fiber of a certain length. Roughly speaking, we will know if a certain ray has generated an exact cover only if its delay (at the $Destination$ node) is equal to the sum of delays induced by all items from $U$.

We know exactly the particular moment when the expected ray (the one which has generated an exact cover) will arrive. This moment is $B$ (the sum of numbers attached to items from $U$). In this case the only thing that we have to do is to "listen" if there is a fluctuation in the intensity of the signal at that particular moment. Due to the special properties of the proposed system we know that no other ray will arrive, at the $Destination$ node, at the moment when the ray generating the exact cover has arrived.

The delays, which are introduced by each arc (which is actually the sum of delays of particular items from $U$), cannot take any values. If we would put random values for delays we might have different rays (which are not covering $U$ exactly) arriving, at the $Destination$ node, in the same time with a ray representing an exact cover. This is why we assign, to each element $u_i$ of $U$, a special number (denoted by $d_i$).

We need only the ray, which has generated the exact cover, to arrive in the $Destination$ node at the moment equal to the sum of delays of each item $u_i$ (the moment when the ray has entered in the $Start$ node is considered moment 0). Thus, the delaying system must have the following property:\\

\textbf{Property of the delaying system}

Let us denote by $d_1$, $d_2$, ..., $d_n$ the delays that we plan to attached to $u_1$, $u_2$, ..., $u_n$. A correct set of values for this system must satisfy the condition:

$d_1+d_2+...+d_n \not= a_1 \cdot d_1+a_2 \cdot d_2+...+a_n \cdot d_n$,

where $a_i$ ($1 \leq i \leq n$) are natural numbers ($a_i \ge 0$) and cannot be all 1 in the same time. 

Basically speaking $a_k$ tell us how many times the ray has passed through an element $u_k$. Thus, if value $a_k$ is strictly greater than 1 it means that the ray has passed at least twice through element $u_k$. Note that an element $u_i$ can belong to multiple subsets from $C$. For instance, in the example from section \ref{xc_example} we can see that element $u_1$ belongs to 3 subsets ($C_1$, $C_2$ and $C_4$). 

Actually each item $u_i$ can appear no more than $m$ times on any path between $Start$ and $Destination$ (see Figure \ref{fig_device3}). This is because the cardinal of $C$ is $m$ and each $C_k$ ($1 \le k \le m$) is a subset of $U$ (that is each element from $U$ can appear no more than once). In our case we were not interested to depend on particular values for $m$. This is why when we have designed the system of numbers $d_i (1 \le i \le n)$ we have assumed that each item $u_i$ can appear any times in the final solution (not limited to $m$ times). This is the most general case. Other cases, where has a fixed value can lead to smaller values for numbers $d_i$.

Some examples of numbers complying with our property are given in Table \ref{tab_labeling_system}.

\begin{table}[htbp]
\begin{center}
\caption{Some examples. First column contains the cardinal of $U$. The second column represents the numbers assigned to each element of $U$.}
\label{tab_labeling_system}
\begin{tabular}
{p{50pt}p{100pt}}
\hline
n& 
Labels (delays) \\
\hline
1& 
1 \\
2& 
2, 3 \\
3& 
4, 6, 7 \\
4& 
8, 12, 14, 15 \\
5& 
16, 24, 28, 30, 31 \\
6& 
32, 48, 56, 60, 62, 63 \\
\hline
\end{tabular}
\end{center}
\end{table}

From Table \ref{tab_labeling_system} it can be easily seen that these numbers follow a general rule. For a problem with $n$ elements in $U$ one of the possible labeling systems is:\\

$2^n-2^{n-1}$, 

$2^n-2^{n-2}$, 

$2^n-2^{n-3}$, 

... ,

$2^n-2^0$. \\

\textbf{Remarks}

\begin{itemize}

\item{In \cite{oltean_uc} these numbers have been used for solving the Hamiltonian path problem with a light based-device. It was proved \cite{oltean_uc} that these numbers satisfy the previously exposed property. In \cite{oltean_nc} it was proved this system is minimal (the greatest number is the smallest possible).}

\item{The numbers in this set are also called Nialpdromes (sequence A023758 from The On-line Encyplopedia of Integer Numbers \cite{sloane}). They are numbers whose digits in base 2 are in nonincreasing order.}

\item{These numbers have been used in \cite{Henkel} for solving NP-complete problems in the context of DNA computers \cite{adleman},}

\item{The delaying system described above has the advantage of being a general one, but it also has a weakness: it is exponential.}

\end{itemize}

\subsection{How the system works}
\label{howorks}

In the graph depicted in Figure \ref{fig_device3} the light will enter in $Start$ node. It will be divided into 2 subrays of smaller intensity. These 2 rays will arrive into the second node at moments $delay(C_1)+k$ and $k$. Each of them will be divided into 2 subrays which will arrive in the $3^{rd}$ node at moments $2*k, delay(C_1)+2*k, delay(C_2)+2*k, delay(C_1)+delay(C_2)+2*k$. These rays will arrive at no more than 4 different moments. 

In the $Destination$ node we will have $2^m$ rays arriving at no more than $2^m$ different moments. The ray arriving at moment $m*k$ means the empty set. The ray arriving at moment $delay(C_1)+delay(C_2)+...+delay(C_n) + m*k$ represents the full set ($C$). If there is a ray arriving at moment $B+m*k$ means that there a subset of $C$ of sum $B$ (this means an exact cover for $U$).

If there are 2 rays arriving at the same moment in the $Destination$ it simply means that there are multiple subsets which have the same sum (generating the same cover - exact or not). This is not a problem for us because we want to answer the YES/NO decision problem (see section \ref{subsetsum}). We are not interested at this moment which is the subset generating the solution.

Because we are working with continuous signal we cannot expect to have discrete output at the $Destination$ node. This means that rays arrival is notified by fluctuations in the intensity of the light. These fluctuations will be transformed, by a photodiode, in fluctuations of the electric signal which is stored into memory and later analyzed.

\section{Complexity}
\label{complexity}

This section answers a very important question: \textit{Why the proposed approach is not a polynomial-time solution for the Exact Cover problem?}

At the first sight one may be tempted to say that the proposed approach provides a solution in polynomial time to any instance of the exact cover problem. The reason behind such claim is given by the ability of the proposed device to provide output to any instance in $O(B)$ complexity. This could mean that we have found a polynomial-time algorithm for the Exact Cover problem. A direct consequence is obtaining solutions, in polynomial time, for all other NP-Complete problems - since there is a polynomial reduction between them \cite{garey}.

However, this is not our case. There are two main reasons for this:

\begin{itemize}

\item{the intensity of the signal decreases exponentially with the number of nodes that are traversed. When the ray is passing through a node it is divided into $2$ subrays. If divided uniformly we could have a $2$ times decrease in the intensity for each subray. If a ray is passing through 10 nodes we can have a decrease of $2^{10}$ times from the initial power. A possible way for handling this issue is discussed in section \ref{amplify}.}

\item{The numbers assigned to each item in $U$ increase exponentially with the cardinal of $U$.}

\end{itemize}

In \cite{Vergis} it was proved that any analog computer can be simulated efficiently (in polynomial time) by a digital computer. From this assumption and the assumption that P $\neq$ NP we can also draw the conclusion that our device is not a polynomial time one.

\section{Physical implementation of the system: difficulties and solutions}

This section describes the components required for the physical implementation of the system and some difficulties that might appear during construction and testing.

\subsection{Components}
\label{hard}

\begin{figure}[htbp]
\centerline{\includegraphics[width=4.9in,height=1.5in]{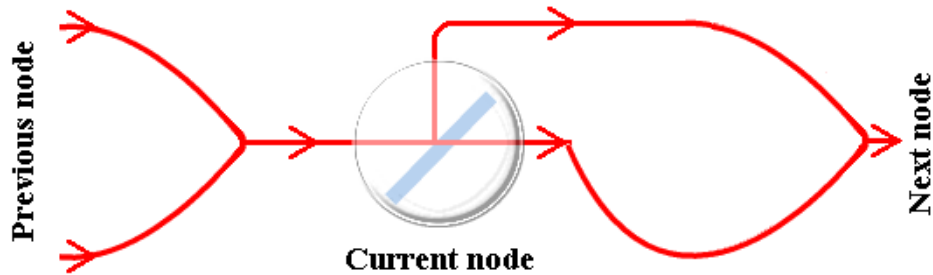}}
\caption{The way in which a ray can be split into 2 sub-rays by using a beam-splitter.}
\label{beam_splitter}
\end{figure}

For implementing the proposed device we need the following components:

\begin{itemize}

\item{a source of light (laser),}

\item{Several beam-splitters for dividing light rays into 2 subrays. A standard beam-splitter is designed using a half-silvered mirror (see Figure \ref{beam_splitter}),}

\item{A high speed photodiode for converting light rays into electrical signals. The photodiode is placed in the $Destination$ node,}

\item{A set of optical fiber cables having lengths equals to the numbers in set $U$ (plus constant $k$) and another set of $m$ cables having fixed length $k$. These cables are used for connecting nodes. }

\end{itemize}

\subsection{Precision}
\label{precision}

Another problem is that we cannot measure the moment $B+m*k$ exactly. We can do this measurement only with a given precision which depends on the tools involved in the experiments. Actually it will depend on the response time of the photodiode.

The rise-time of the best photodiode available on the market is in the range of picoseconds ($10^{-12}$ seconds). This means that if a signal arrives at the $Destination$ in the interval $[B+m*k-10^{-12}, B+m*k+10^{-12}]$ we cannot be perfectly sure that we have an exact cover or another one which does not have the wanted property. This problem can be avoided if all cables are long enough. In what follows we will try to compute the length of the cables.

We know that the speed of light in vacuum is $3 \cdot 10^{8} m/s$. Based on that we can easily compute the minimal cable length that should be traversed by the ray in order to be delayed with $10^{-12}$ seconds. This is obviously 0.0003 meters. This value is the minimal delay that should be introduced by an arc.

Please note that the speed of light in different other media is smaller than the speed of light in vacuum (due to the refractive indices). This can help us to solve larger instances of the problem. Table \ref{tab_refractive} shows the refractive indices of different materials and the minimal cable length required in order to induce a delay of $10^{-12}$.

\begin{table}[htbp]
\begin{center}
\caption{The minimal cable length required to induce the $10^{-12}$ seconds delay (fourth column). The cable is made of materials in the first column. The refractive indices are given in second column ( frequency=5.09x1014 Hz). The speed of light of light in that material is given in the third column. Data are taken from \cite{refractive_wiki}}
\label{tab_refractive}
\begin{tabular}
{p{50pt}p{50pt}p{50pt}p{50pt}}
\hline
Material& 
Refractive index &
Speed of light (m/s) &
Cable length (meters)\\
\hline
Vacuum& 
1&
300000000 &
0.0003 \\
Water Ice& 
1.544& 
194300518 &
0.00019\\
Diamond& 
2.419&
124018189&
0.00012 \\
Silicon& 
4.01&
74812967&
0.00007 \\
\hline
\end{tabular}
\end{center}
\end{table}

In what follows we consider the speed of light in vacuum. 

Note that all lengths must be integer multiples of 0.0003. We cannot allow to have cables whose lengths can be written as $p*0.0003 + q$, where $p$ is an integer and $q$ is a positive real number less than 0.0003 because by combining this kind of numbers we can have a signal in the above mentioned interval and that signal does not encode a subset whose sum is the expected one.

Once we have the length for that minimal delay is quite easy to compute the length of the other cables that are used in order to induce a certain delay. First of all we have to divide all delays introduced by $C_k (1 \le k \le m)$ with such factor that the less significant digit (greater than 0) to be on the first position before the decimal place. For instance if we have $delay(C_1) = 100$ and $delay(C_2) = 2000$ we have to divide both numbers by 100. 

After this operation we will multiply the obtained numbers by 0.0003 factor.

This will ensure that if a signal will arrive in the interval $[B+m*k-10^{-12}, B+m*k+10^{-12}]$ we can be sure that it encodes the sum $B+m*k$.

\subsection{Problem size}
\label{psize}

We are interested in computing the size of the instances that can be solved by using a limited amount of resources (cables).

Recall from section \ref{precision} that the minimal delay is 0.0003 m.

This value is the minimal delay that should be introduced by an arc in order to ensure that the difference between the moments when two consecutive signals arrive at the destination node is greater or equal to the measurable unit of $10^{-12}$ seconds. This will also ensure that we will be able to correctly identify whether the signal has arrived in the destination node at a moment equal to the sum of delays introduced by each item in $U$. No other signals will arrive within a range of $10^{-12}$ seconds around that particular moment.

Once we have the length for that minimal delay is quite easy to compute the length of the other cables that are used in order to induce a certain delay.

Recall from section \ref{labeling}, Table \ref{tab_labeling_system} that an instance with 5 items has the following delaying system:\\

16, 24, 28, 30, 31.\\

From the previous reasoning line we have deduced that the smaller indivisible unit is 0.0003. So, we have to multiply these numbers by 0.0003. We obtain: \\

0.0048, 0.0072, 0.0084, 0.009, 0.0093.\\

These numbers represent the length of the cables that must be used in graph's arcs in order to induce a certain delay.

Assume that we have available for our experiments some cables from the internet networks. They can be easily used for our purpose. Assuming that the longest cable that we have is about 300 kilometers we may solve instances with about 28 items in set $U$ \footnote{Assume that each set in $C$ has no more than 1 element. Otherwise the cables' length for each element in a particular set $C_i$ should be summed up.}. This value was infered from equation $2^n * 0.0003m = 300km$.

Note that the maximal number of items in $U$ can be increased when the precision of our measurement instruments (photodiode) is increased.

\subsection{Power decrease}
\label{amplify}

Beam splitters are used in our approach for dividing a ray in two subrays. Because of that, the intensity of the signal is decreasing. In the worst case we have an exponential decrease of the intensity. For instance, in a graph with $m + 1$ nodes ($Destination$ node is not counted because there is no split there), each signal is divided (within each node) into 2 signals. Roughly speaking, the intensity of the signal will decrease $2^m$ times.

This means that, at the $Destination$ node, we have to be able to detect very small fluctuations in the intensity of the signal. For this purpose we can use a photomultiplier \cite{Flyckt} which is an extremely sensitive detector of light in the ultraviolet, visible and near infrared range. This detector multiplies the signal produced by incident light by as much as $10^8$, from which even single photons can be detected.

Another way to amplify the signal is to use doped optical fiber as a gain medium \cite{Desurvire,Mears}. For this operation the signal to be amplified and a pump laser are multiplexed into the doped fiber. Thus, the signal is amplified through interaction with the doping ions. The costs for this operation could be high because the amplification might be needed after each node of the proposed device. More than that, the noise introduced through Amplified Spontaneous Emission might generate unwanted fluctuations in the intensity of the signal.

Also note that this difficulty is not specific to our system only. Other major unconventional computation paradigms, trying to solve NP-complete problems share the same fate. For instance, a quantity of DNA equal to the mass of Earth is required to solve Hamiltonian Path Problem instances of 200 cities using DNA computers \cite{Hartmanis}.

\subsection{Technical challenges}
\label{tech}

There are many technical challenges that must be solved when implementing the proposed device. Some of them are:

\begin{itemize}

\item{Cutting the optic fibers to an exact length with high precision. Failing to accomplish this task can lead to errors in detecting if there was a fluctuation in the intensity at moment $B+m*k$,}

\item{Finding a high precision photodiode. This is an essential step for measuring the moment $B+m*k$ with high precision (see section \ref{precision}).}

\end{itemize}

\section{Improving the device}
\label{speed_reduce}

The speed of the light in optic fibers is an important parameter in our device. The problem is that the light is too fast for our measurement tools. We have either to increase the precision of our measurement tools or to decrease the speed of light.

It is known that the speed of light traversing a cable is significantly smaller than the speed of light in the void space. Commercially available cables have limit the speed of the ray wave up to $60\%$ from the original speed of light. This means that we can obtain the same delay by using a shorter cable (see Table \ref{tab_refractive}).

However, this method for reducing the speed of light is not enough for our purpose. The order of magnitude is still the same. This is why we have the search for other methods for reducing that speed. A very interesting solution was proposed in \cite{Hau} which is able to reduce the speed of light by 7 orders of magnitude and even to stop it \cite{Bajcsy,Liu}. In \cite{Bajcsy} they succeeded in completely halting light by directing it into a mass of hot rubidium gas, the atoms of which, behaved like tiny mirrors, due to an interference pattern in two control beams.

This could help our mechanism significantly. However, how to use this idea for our device is still an open question because of the complex equipment involved in those experiments \cite{Hau,Liu}.

By reducing the speed of light by 7 orders of magnitude we can reduce the size of the involved cables by a similar order (assuming that the precision of the measurement tools is still the same). This will help us to solve larger instances of the problem.

\section{Conclusions and further work}
\label{further}

The way in which light can be used for performing useful computations has been suggested in this paper. The techniques are based on the massive parallelism of the light ray.

It has been shown the way in which an optical device can be used for solving the Exact Cover problem.

Further work directions will be focused on:

\begin{itemize}

\item{Implementing the proposed device,}

\item{Automate the construction process. Cutting new cables each time a new instance has to be solved is very inefficient. This is why finding a way to reuse the exiting cables (from previous instances) is a priority for our research,}

\item{Our optical solution cannot output the exact cover representing the solution. It can only say if there is a solution or not. However, the YES/NO decision variant of the XC problem is still NP-complete \cite{garey}. We are currently investigating a way to store the visited arcs so that we can easily reconstruct the path,}

\item{Using Microelectromechanical Mirror Array \cite{Rader} for inducing delays,}

\item{Finding other non-trivial problems which can be solved by using the proposed device,}

\item{Finding other ways to introduce delays in the system. The current solution requires cables that are too long and too expensive,}

\item{Using other type of signals instead of light. Possible candidates are electric power and sound.}

\end{itemize}

\begin{acknowledgment}

The author likes to thanks to the anonymous reviewers for providing useful suggestions on an earlier version of this paper \cite{oltean_uc}.

This work was supported by grants UBB-TP\_T-30945/2007 and CNCSIS-IDEI-543/2007. 

\end{acknowledgment}

\end{document}